\documentclass[twocolumn]{revtex4}

\def\be{\begin{equation}} 
\def\ee{\end{equation}}
\def\bea{\begin{eqnarray}} 
\def\eea{\end{eqnarray}}
\def \line{\hbox to \hsize}    
\def\frac #1#2{{#1\over #2}}

\def\psid{\psi^{\dagger}}

\def\Psid{\Psi^{\dagger}}

\def \ket #1{{\vert #1\rangle}}

\def\1{\mbox{\bf 1}}
\newcommand{\comment}[1]{}

\newcommand{\details}[1]{}

\def \mh{\mathcal{H}}
\newcommand{\para}[1]{}
\usepackage{bm}
\usepackage{hyperref,amsmath,amssymb,bm}

\usepackage{color,epsf,graphicx,amssymb,longtable,afterpage,mathrsfs}

\date{13 January 2010}

\begin{document}
\title{Topological Majorana and Dirac zero modes in superconducting vortex cores}
\author{Rahul Roy}

\affiliation{Rudolph Peierls Centre for Theoretical Physics,\\
1 Keble Road, Oxford, OX1 3NP, UK}

\begin{abstract} 

We provide an argument based on flux insertion to show that certain
superconductors with a non-trivial topological invariant have
protected zero modes in their vortex cores. This argument has the
flavor of a two dimensional index theorem and applies to disordered
systems as well. It also provides a new way of understanding the zero
modes in the vortex cores of a spinless $p_{x} + i p_{y}$
superconductor. Applying this approach to superconductors with and
without time reversal and spin rotational symmetry, we predict the
necessary and sufficient conditions for protected zero modes to exist
in their vortices.

\end{abstract}
\maketitle

\details{Introduction}
\label{sec-1}
\details{Introduction}
\label{sec-1.1}

  Fermionic zero energy modes are predicted to exist in defects in a
variety of condensed matter systems from one dimensional (1d) chains
of polyacetylene to the vortices of certain two dimensional (2d)
superconductors \cite{wpsu80a,kopnin91a}. These modes typically occur
in systems with a bulk gap which are in a topologically non-trivial
phase. The zero energy modes are often topologically protected in the
sense that their existence is only tied to the particular topological
phase of the system and to some global properties of the system such
as the winding number of the phase of the order parameter in a
superconductor.

  Unpaired fermionic zero energy modes which are localized in the
vortices of certain superconductors such as a spinless $p_{x} + i p_{y}$
superconductor are of special interest. These modes, also called
Majorana modes, have attracted a fair bit of recent theoretical
attention due to the fact that they can obey non-abelian statistics
and have potential applications in decoherence free quantum
computation \cite{kitaev97a,read00a,ivanov01a,nayak07a}. The existence
of these modes has traditionally been demonstrated through explicit
analytic solutions of the Bogoliubov de-Gennes (BdG) equations
\cite{stone06a,stone04a,read00a,gurarie07a,kopnin91a}.

  In 1d, Jackiw and Rebbi showed that the Dirac equation interacting
with a scalar field with a topologically interesting configuration
such as a soliton has zero energy solutions whose number was related
to an integer which characterizes the non-trivial topology
\cite{jackiw76a}.  Similar zero energy states also occur in
polyacetylene \cite{wpsu80a}. Recently many works have attempted to explain 
the Majorana modes in 2d by reducing the problem to a 1d problem 
\cite{stewari07a,dhlee07a,kitaev06a}.

  Since the vortex zero energy states occur in 2d systems, a more
general and inherently 2d explanation of the Majorana zero modes is
desirable. In the current work, this is achieved through a flux
insertion argument similar to Laughlin's gauge argument
\cite{laughlin81a}.  The argument is not tied to any particular model
and also applies to disordered systems where translational invariance
is broken and is particularly useful in the context of recent efforts
to find Majorana modes in systems other than a $p_{x}+ ip_{y}$
superconductor \cite{jdsau09a,sato09a,alicea09a,plee09a}. One of our
main results is a prediction about precisely what systems in all the
different symmetry classes of the classification introduced by Altland
and Zirnbauer\cite{altland97a} support zero energy states.

 Using a flux insertion argument, we first show that a certain class
of insulators with \(\pi\) flux inserted through a plaquette have
exact zero energy eigenstates which are topologically protected in the
sense that their existence is connected to the Hall conductance of the
insulator.  We then use these results to study superconductors with
and without time reversal and spin rotational symmetry and predict
precisely which systems have protected zero modes.
\details{Insulators}
\label{sec-2}
\details{\details{insulators 1}}
\label{sec-2.1}

 The insulator Hamiltonians studied below have the same symmetries as
the mean field Bogoliubov-de-Gennes Hamiltonians of
superconductors. Though such insulators are possibly of little
physical significance by themselves, we study them since their spectra
are identical to that of the BdG Hamiltonians of
superconductors. Further since they are insulators rather than
superconductors, they do not expel flux, making a flux insertion
argument possible. Consider a 2d tightbinding insulator on a lattice
with an even number of orbitals, $2s$ per site.  The basis states are
written in the form $\ket{i,\alpha}$ where $i$ is an index for the
position and $\alpha$ for the orbital. We first study infinite sized
systems where the energy of the topologically protected vortex states
is exactly zero and later consider corrections that would occur in a
finite sized system.
\details{\details{insulators 2}}
\label{sec-2.2}

The Hamiltonian can be written in the form
 \(
\mh = \sum_{i,j} \Psid_i H_{ij}\Psi_j
\)
where $H_{ij}$ is a $2s\times 2s$ dimensional matrix and \( \Psi_j =
(\psi_{j,\alpha})^{T} ,\Psid_i = (\psid_{j,\beta}) \) where $i,j$
are the lattice position indices and $\alpha,\beta$ the orbital and spin
indices. The Fermi energy is set to zero and is assumed to lie in a gap.

  In the first instance, we study systems with neither spin rotational
nor time reversal symmetry, though we will analyze systems with these
symmetries later on.  We further restrict our study to Hamiltonians
$\mh$ which posses a symmetry analogous to that of Bogoliubov
de-Gennes Hamiltonians.  In other words, we assume the existence of an
anti-unitary operator, $S$, such that
\begin{equation}
\label{eq:4}
S\mh S^{-1}=-\mh.
\end{equation} 
Further, we assume that $S$ acting on  single particle position eigenkets in the
Hilbert space produces
 a linear combination of kets at the same position, i.e., \(S \Psi_j
 S^{-1}= U_{j}\Psi_j\) and $U_j$ is a $2s\times 2s$
 dimensional unitary matrix.  We may therefore conclude that the
 Hamiltonian of the system despite being that of an insulator, belongs
 to the symmetry class D in the classification introduced by Altland
 and Zirnbauer\cite{altland97a}. The symmetry under $S$ also implies
 that $U_{i} H_{ij} U_{j} ^{-1}= -H_{ij}^{*}$.
\details{\details{insulators contd}}
\label{sec-2.3}

 Consider the effect of a vector potential which arises from flux
insertion through an infinitesimal tube. The Hamiltonian of the system
with a flux tube somewhere in the sample may be derived by multiplying
each hopping term in the matrix $H_{ij}$ by the phase factors $\exp{[
i(e/\hbar) \int_{j} ^{i} \bm{dr}.~\bm{A}]}$
 where the integral is
along the hopping path and $\bm{A}$ is the vector potential. The
Hamiltonian of the system with the vector potential, which we call
\(\mh(\bm{A})\) then has the following property:
\begin{eqnarray}
  S\mh(\bm{A}) S^{-1}&=& S(\sum_{i,j} \Psid_i H_{ij}\Psi_j e^{ i(e/\hbar) \int_{j} ^{i} \bm{dr}.~\bm{A}} ) S^{-1} \\ \nonumber
  &=& -\sum_{i,j}\Psid_i H_{ij} \Psi_j
  e^{- i (e/\hbar)\int_{j} ^{i} \bm{dr}.~\bm{A}} \\ \nonumber
  &=& - \mh(-\bm{A}) \nonumber.
\end{eqnarray}
\details{\details{flux insertion argument}}
\label{sec-2.4}

Consider an infinite sample of the above system and let us
adiabatically thread flux through an infinitesimal flux tube through a
plaquette at the center of the sample.
 If $\mh(\phi)$ is the Hamiltonian in the presence of a flux $\phi$,
then on the basis of the above, we conclude that
\begin{equation}
  \label{1}
  S\mh(\phi)S^{-1} = - \mh(-\phi).
\end{equation}
 Thus if $\mh(\phi)$ has an eigenstate with eigenvalue $E$, then
$\mh(-\phi)$ has an eigenstate with eigenvalue $-E$.
\details{\details{insulators}}
\label{sec-2.5}

   Now suppose the system has a quantized Hall conductance of
$pe^2/2\pi\hbar$. Then as a flux of $2\pi\hbar/e$ is adiabatically
inserted through the flux tube, a total charge of $pe$ flows in from
infinity towards the flux tube \cite{laughlin81a}. In an infinite
sample, the total spectral flow, i.e., the total number of states
which cross the gap at the Fermi surface from below minus the number
of states which cross it from above is equal to $p$
\cite{laughlin81a,halperin82a}.
\details{\details{insulators contd}}
\label{sec-2.6}

Let $n(\phi)$ be the number of eigenstates of the Hamiltonian whose
 energy is zero at the flux $\phi$. 
A schematic example of the energy spectrum of extended states close to
the Fermi energy as a function of flux is plotted
in Fig. 1.  The function $n(\phi)$ is non-zero only for \(\phi \in
\{\phi_a, \phi_b, {2\pi\hbar\over 2e} ,\phi_{c}, \phi_d\} \) where 
\(\phi_{c}= 2\pi -\phi_{a} \) and \( \phi_{d}= 2\pi-\phi_{b}\).

\begin{figure}[b] 
\includegraphics[width=0.45\textwidth]{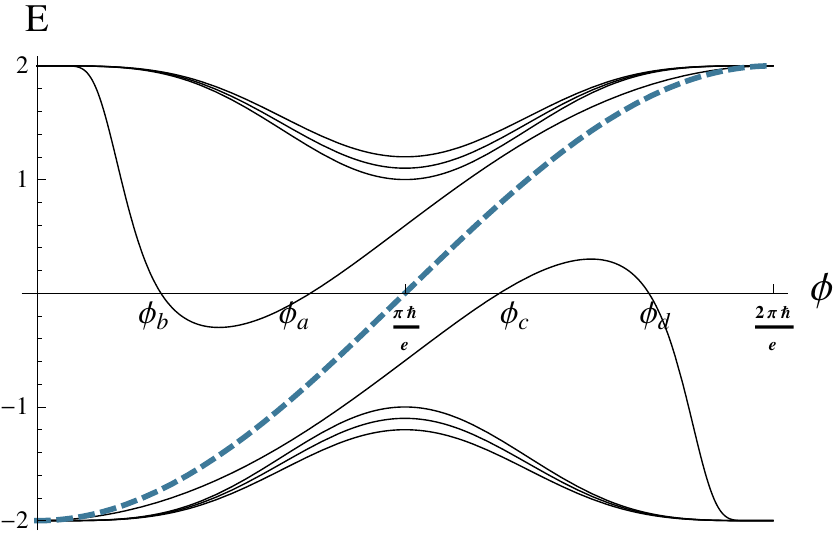}
\caption{A schematic plot of the energy, $E$ versus  $\phi$,
where $\phi$ is the flux inserted through a plaquette
of the insulator with Bogoliubov symmetry. Only states which lie close to the 
Fermi energy at $\phi=0$ are shown. The dashed line is the single energy
curve which traverses the gap as the flux is changed from 0 to $2\pi\hbar/e$.
 }  \label{fig2}
\end{figure}

From Eq.~\eqref{1},
\begin{equation}
  \label{2}
  n(2\pi- \phi) = n(\phi).
\end{equation}

Apart from the states which traverse the gap and thus cross 
the Fermi energy an odd number of times, there might
also be states which cross the Fermi energy  an even number of 
times as shown in the figure. 

Thus, 
 \begin{equation}
 \label{3}
  \sum_{\phi_i \ni n(\phi_{i})\ne 0} n(\phi_i) = p+2m,
\end{equation}
 where $p$  is the total Chern number of the ground state.

 It follows from Eqs.~\eqref{2} and \eqref{3} that when $p$ is odd, 
\[
 n(0) + n(2\pi\hbar/2e) = 2k+1
\]
where $k$ is some integer. 
\details{\details{insulators contd}}
\label{sec-2.7}

 At zero flux, the Fermi energy lies in a gap.  
Thus, $n(0)=0$ which in turn implies that $n(2\pi\hbar/2e)$ is an odd integer
when $p$ is odd. Since the integer $p$ is a topological invariant, which cannot
change under small transformations of the Hamiltonian, the zero mode is
topologically protected. 
\details{Superconductors}
\label{sec-3}
\details{\details{sc1}}
\label{sec-3.1}

 We now use this result to study superconductors.  In the remainder of the paper,
we are frequently going to consider Hamiltonians with $2s$ degrees of freedom 
per site and we write such matrices in the form:
\begin{equation}
\label{eq:2}
\begin{pmatrix}
M_{11} & M_{12} \\
M_{12}^{\dagger} & M_{22}
\end{pmatrix}.
\end{equation}

Consider a tightbinding BCS Hamiltonian for fermions hopping on a
lattice which can be written in the form: \( \mh_{S} = \sum_{i,j}
\Psi^{N\dagger}_{i}H^{B}_{ij}\Psi^{N}_{j} \) where \(\Psi_i ^{N} =
(\varphi_{i,\gamma},\, \varphi^{\dagger}_{i,\gamma})^{T} \) is a Nambu
spinor. Here, $i,j$ stand for positions, and $\gamma$ is an index for
the orbitals and spin which runs from 1 to $s$.  The matrix $H^{B}$
is the BdG Hamiltonian which is of the form of Eq.~(\ref{eq:2}) with
\(M_{11}=h, M_{12}=\Delta \) and \( M_{22}= -h^{T}\) where ${h}$ is
the single particle Hamiltonian and ${\Delta}$ is the gap matrix.  We
can map the Hamiltonian $\mh_{S}$, to the Hamiltonian of an insulator,
$\mh_{I}$ (which we call the associated insulator) given by \( \mh_{I}
= \sum_{ij}\Psid_{i} H^{B}_{ij}\Psi_{j} \) where \(\Psi_{j}=
(\psi_{i,\alpha},\,\psi_{i,\beta})^{T}\).

 We first study BdG Hamiltonians with neither time reversal nor spin
rotational symmetry. Let us imagine inserting an infinitesimal flux
tube containing a flux of $2\pi\hbar/2e$ at the origin of our
coordinates which is placed at the center of one of the plaquettes of
the system.  We imagine that this flux does not leak into the
rest of the superconductor and that the flux is inserted in such a way
that the low energy configuration where the phase of the
superconductor winds by $2\pi$ around the flux tube is attained.

 In the presence of the flux, the single particle Hamiltonian
gets transformed as follows. Let $r_i,\, \theta_i$ be respectively the distance from the origin
and the polar angle of site $i$ in a coordinate system with the
origin located at the position of the flux tube. Then,
 \( h_{ij} \rightarrow h'_{ij} = h_{ij}
e^{i{(e/\hbar)}\int\bm{dr}.\bm{A}} =
h_{ij}e^{i(\theta_i-\theta_j)/2}\), where $\bm{A}$ is chosen to be \(
{\hbar \over 2e} \nabla(\theta)\). The gap matrix transforms as: \(
\Delta_{ij} \rightarrow \Delta' _{ij}= \Delta e^{i(\theta_i +
\theta_j)/2}\) in the presence of the flux tube.
\details{\details{version 2}}
\label{sec-3.2}

 The BdG eigenvalue equation in the presence of the vortex is thus: 
\(
H' \psi= E \psi
\), where \(\psi= ( u,\,v)^{T}\) and  $H'$ has the form of
Eq.~(\ref{eq:2}) with $M_{11}=h', M_{12}=\Delta'$ and $M_{22}=-(h')^{T}$.  Let \(
\tilde{u}_i = u_i, \tilde{v}_i = e^{i\theta_i } v_i \) Then
\(\tilde{\psi}=( \tilde{u},\, \tilde{v} )^{T} \) satisfies the
eigenvalue equation : 
\( H'' \tilde{\psi} = E \tilde{\psi}
\), where $H''$ can be written in the form of Eq.~(\ref{eq:2}) with 
\(M_{11}=h', M_{12}=\Delta'', M_{22}=-(h''), h''_{ij}=
h_{ji} e^{i(\theta_i -\theta_j)/2}\) and $\Delta''_{ij} = \Delta_{ij}
e^{i(\theta_i - \theta_j)/2}$.
\details{\details{ superconductors contd}}
\label{sec-3.3}

 We now replace the superconductor with the associated 
insulator, $\mh_{I}$ with the
same flux configuration, i.e., half a quantum of flux inserted at the origin,
which lies at the center of a plaquette. It is easy to verify
that  $\mh_{I}(\pi)$ can be written as $\sum_{ij}\Psid_{i} H''_{ij} \Psi_{j}$ with 
\(H''\) as given above. 
Further, $\mh_{I}$ satisfies the symmetry in Eq.~(\ref{eq:4}) since it is 
derived from a BdG Hamiltonian and the analysis for the zero modes for
insulators made previously can therefore be used. 

The necessary and sufficient condition for the existence of an exact
zero-energy mode which is localized around the flux tube is therefore
that the Hall conductance of $\mh_{I}$ is $pe^{2}/2\pi\hbar$ where $p$ is odd.
If this condition is met, it follows that there is a zero energy mode
localized around the plaquette containing the tube. Since the Hall conductance
is a robust topological invariant, the existence of the zero mode 
for the superconductor is also topologically protected.  

 A more realistic description of a vortex would include a finite
region larger than a single plaquette where the magnetic field is
non-zero rather than the situation considered above where the flux is 
confined to a single plaquette. 
 In the general case, the gap parameter and the single
particle Hamiltonians may be modeled as:
\( 
\Delta'_{ij}= \Delta_{ij}f((r_i + r_j)/2)e^{i(\theta_i + \theta_j)/2};
 h'_{ij} = h_{ij}e^{i \phi_{ij}} 
\). Here, $f(r)$ is some function which is zero at the origin and goes to
one for distances larger than the coherence length and $\phi_{ij} =
\int (e/\hbar)\bm{dr.A}$ where $\bm{A}$ is the vector potential which corresponds
to a smeared out flux and which approaches the value \( (\hbar/e)
\nabla (\theta)/2\) 
for distances much larger than the penetration depth.

The corresponding BdG Hamiltonian may therefore be written as
$\mh=\mh_0 + V$ where $\mh_0$ is the Hamiltonian which corresponds to
the idealized vortex discussed above and $V$ is a local perturbation
in the sense that \(V_{ij} = 0 \) sufficiently far from the vortex.
If there is an odd number of Majorana zero energy states, then the
Hilbert space of the idealized vortex has one unpaired fermionic
mode, while every other eigenstate of energy, $E$, can be paired with
a state of energy, $-E$.
 Any local perturbation cannot alter the Hilbert space
structure. It follows that the perturbation, $V$, does not alter the
existence of a zero energy mode and that the zero mode persists in the
more realistic configuration.
\details{\details{branch2}}
\label{sec-3.4}

 We now intend to study superconductors with time reversal symmetry
using the same technique and therefore first study the corresponding
insulators.  Insulators with time reversal symmetry are classified by
a $Z_{2}$ invariant. Any Hamiltonian, $\mh$, of this class can be
continuously deformed, without closing the gap or breaking time reversal
symmetry, to a Hamiltonian
$\mh'$ which is diagonal in the basis of the z-component of spin and
can thus be written in the form \( \mh' = \mh'_{\uparrow} +
\mh'_{\downarrow} \).  The Hamiltonians $\mh'_{\uparrow}$ and
$\mh'_{\downarrow}$ can be analyzed separately for studying their zero
modes. The sum of the Hall conductance of $\mh'_{\uparrow}$ and
$\mh'_{\downarrow}$ must be zero. If $|\sigma_{xy}2\pi\hbar/e^{2}|$ of
$\mh'_{\uparrow}$ is an odd number, then the insulator belongs to the
non-trivial class. By the previous analysis, in this case, when a flux
of $\pi$ is inserted through the central plaquette the system has an
odd number of Kramers pairs at zero energy, one member of each pair
associated with $\mh'_{\uparrow}$ and the other with
$\mh'_{\downarrow}$. Kramers theorem then prevents a gap from opening
up at $\pi$ flux when the Hamiltonian is continuously deformed without
breaking time reversal symmetry. This implies that the original
Hamiltonian, $\mh$, must also have an odd number of pairs of zero
modes at $\pi$ flux. When $\sigma_{xy}(\mh_{\uparrow}')$ is an even
multiple of $e^{2}/2\pi\hbar$ on the other hand, the number of pairs of zero
modes must be even and is therefore not protected.
\details{\details{time reversal etc.}}
\label{sec-3.5}

Superconductors with time reversal symmetry fall in the class DIII of
the classification scheme \cite{altland97a}. These superconductors are
classified by a $Z_{2}$ invariant \cite{roy06a} and can be mapped onto
insulators with time reversal symmetry which satisfy Eq.(\ref{eq:4})
and which have the same spectrum, exactly as in the case of
superconductors without time reversal or spin rotational symmetry.  We
have shown in the previous paragraph that when flux is inserted
through a plaquette in these associated insulators, they have an odd number of
Kramers pairs of zero energy modes if and only if they belong to the
non-trivial topological class. The zero eigenmodes of the equation
that determines the spectra of the insulator with half a quantum of
flux inserted through a plaquette may be related to the zero
eigenmodes of the superconductor with a vortex following the
discussion after Eq.~(\ref{eq:2}). Thus, superconductors with time
reversal symmetry have an odd number of Kramers pairs of zero energy
modes in their vortex cores whenever the superconductors are in the
non-trivial topological class. A pair of zero-energy modes may be
regarded as a single Dirac mode.
\details{\details{version 2}}
\label{sec-3.6}

  Hamiltonians of superconductors with spin-rotational symmetry but
not time reversal symmetry fall in the class C of the Altland-Zirnbauer 
symmetry
classes \cite{altland97a}. These Hamiltonians may be regarded as the sum of
\(
 \mh_{\uparrow}=(\psid_{\uparrow},\,
\psi_{\downarrow}) H_{\uparrow}  (\psi_{\uparrow},\,\psid_{\downarrow} )^{T}\)
and \(\mh_{\downarrow}= (\psid_{\downarrow},\,
\psi_{\uparrow}) H_{\downarrow}  (\psi_{\downarrow},\,\psid_{\uparrow} )^{T}
\)\cite{altland97a}, where $H_{\uparrow}$ and $H_{\downarrow}$ have the form of 
Eq.~(\ref{eq:2}) with \(M_{11} = h, M_{12}=\Delta, M_{22}=-h^{T}\) and 
\(M_{11}= h,M_{12}=-\Delta, M_{22}=-h^{T}\) respectively. 
\details{Table and summary}
\label{sec-4}

\begin{table}[htb]
\caption{\label{tbl:results}Conditions  for  superconductors in the various symmetry classes to support protected zero modes in vortex cores, expressed as conditions on $\mh_{I}$, the insulator associated with the superconductor. The last column indicates whether there is a single protected Majorana mode(M) or a protected pair of modes(D). No protected modes exist for the class CI.}
\begin{center}
\begin{tabular}{|l|l|l|l|l|l|}
\hline
 Class  &  Time-rev  &  Spin-rot  &  Condn. on $\mh_{I}$                     &  Mode  \\
\hline
 D      &  No        &  No        &  \(\sigma_{xy}2\pi\hbar/e^{2}=2k-1\)     &  M     \\
\hline
 C      &  No        &  Yes       &  \(\sigma_{xy}2\pi\hbar/e^{2}=2(2k-1)\)  &  D     \\
\hline
 DIII   &  Yes       &  No        &  non-trivial $Z_2$                       &  D     \\
\hline
 CI     &  Yes       &  Yes       &  -                                       &  -     \\
\hline
\end{tabular}
\end{center}
\end{table}

 The spectra of $H_{\uparrow}$ and $H_{\downarrow}$ are identical. If
 \( (u,v)^{T} \) written in the particle hole basis, is a zero mode of
 $H_{\uparrow}$, then \( (u,-v)^{T} \) is a zero mode of $H_{\downarrow}$.
 The superconductor may be mapped onto an insulator, $\mh_{I}$, which
 is the sum of two single particle Hamiltonians, $\mh_{I,\uparrow}$
 and $\mh_{I,\downarrow}$, and which may be regarded as separate
 systems.  The condition that in the presence of a vortex, the matrix
 $H_{\uparrow}$ has an odd number of zero modes is, as deduced in the study
of Hamiltonians of class D,
 that the Hall conductance of the corresponding insulator,
 \(\sigma_{xy}(\mh_{I,\uparrow})2\pi\hbar/e^{2}\) is an odd integer. The net Hall
 conductance, $\sigma_{xy}(H_{I})$ is twice that of $H_{I,\uparrow}$ and
 is thus always an even integer. Thus, when the Hall conductance
 of $\mh_{I}$ has the form $2pe^{2}/2\pi\hbar$ where $p$ is an odd integer, the system
 has an odd number of pairs of zero modes in its vortices, while when
 $p$ is even, the system has an even (or zero) number of pairs of zero
 modes.  

Superconductors with both time reversal and spin rotational
 symmetry, which belong to the class CI, may be regarded as belonging
 to the trivial $Z_{2}$ class of superconductors with time reversal
 symmetry. These superconductors thus have no topologically protected
 zero modes in their vortex cores.

 Our results are summarized in Table \ref{tbl:results}. In the
  notation used in Ref.~\cite{schnyder08a}, superconductors in the
  class D have protected Majorana modes in their vortex cores when the
  integer invariant for this class is odd, superconductors in class C
  and DIII have protected Dirac modes when the invariants for these
  classes are odd and non-trivial respectively.  Most continuum models
  can be simulated to an arbitrary degree of accuracy by a series of
  lattice models. Since the results derived above are not limited to a
  particular tightbinding model, one expects that the analysis
  presented above extends also to continuum models \footnote{The
  exception is the class of models, for which there are topological
  obstructions to simulation on a lattice such as models with time
  reversal symmetry and a single Dirac cone as the model in
  Ref.~\cite{fu08a}}. 

  Our analysis thus far has been restricted to a single vortex in an
  infinite superconductor. A finite superconductor with a vortex is
  topologically equivalent to a system with two vortices and periodic
  boundary conditions.  One may then start with the associated
  insulator with no flux inserted and gradually insert flux into two
  tubes in such a way that the flux entering one tube is equal to the
  flux leaving the other leaving the total flux through the system
  zero. At $2\pi\hbar/2e$ flux, the states localized at the two flux
  tubes will in general hybridize, giving rise to a finite
  splitting. The magnitude of the splitting when the two vortices are a
  certain distance $d$ is proportional to the overlap between two
  zero-energy eigenstates of the infinite system placed at the same distance and
  therefore falls exponentially as the distance between the vortices.

 Finally, we note that the arguments made above would are also
applicable in the case when there is a mobility gap in the absence of flux
rather than a gap to all states since the states within the mobility
gap have zero Chern number and their energies return to their initial
values when the flux inserted varies from $0$ to $2\pi\hbar/e$.

In summary, we have provided a
simple and general argument which shows that certain topological
classes of superconductors have topologically protected, robust zero
modes, which can either be unpaired Majorana modes, or come in pairs.
We applied this analysis to the various symmetry classes of
superconducting Hamiltonians.

 The author is grateful to Steven Simon and John Chalker for extensive
discussions and inputs which have helped shape the current manuscript
to its final form. The author also gratefully acknowledges support
from EPSRC grant EP/D050952/1.
\details{Bibliography}
\label{sec-5}

\end{document}